\documentstyle[12pt,bezier,emlines2]{article}
\newcommand{\tr}[1]{\,{\rm tr}\,#1\,}
\begin{document}
\title{
\begin{flushright}
{\small SMI/TH-18-96 }
\end{flushright}
\vspace{2cm}
Planar Approximation as Two-Field Boltzmann Theory}
\author{
I.Ya.Aref'eva
\thanks
{ e-mail: arefeva@arevol.mian.su}
\\
Steklov Mathematical Institute,\\
Vavilov 42, GSP-1, 117966, Moscow, Russia}
\date {$~$}
\maketitle
\begin {abstract}
A modified  interaction representation for the master field
describing connected $SU(N)$-invariant Wightman's functions
in the large $N$ limit of matrix fields is constructed.
This construction is based on the representation of the master
field in terms of Boltzmannian field theory found before \cite {AVM}.
In the modified  interaction representation we deal with two
scalar Boltzmann  fields ({\it up} and {\it down} fields).
For up and down fields only half-planar diagrams contribute and
this could help to write down a recursive set of non-linear
integral-differential equations summing up planar diagrams.
\end {abstract}


Recently the master field describing the large N limit in matrix models
and QCD in four dimensional space-time has been constructed \cite{AVM}.
The obtained master field satisfies to the ordinary equations of
relativistic field theory but fields are quantized according
to a new rule in so called Boltzmannian Fock space.
Boltzmann field acting in the free (Boltzmannian) Fock space was
used by Gopakumar and Gross \cite {GG} and
Douglas \cite {Doug} in construction of the master field for 0-dimensional
matrix  models (for discussion of the Boltzmannian Fock space
see also \cite {ALV} and refs. therein). Boltzmann field was also used in
construction of the master field for QCD$_{2}$
\cite {Gopak}.
About definition of the master field and early attempts of construction
of the master field see refs. in \cite{AVM,Aluk}.

The construction  \cite{AVM} of master field is based on 
the Yang-Feldman equations
and it deals not with
Green's functions but with  the Wightman's correlation functions.
As it is well known in the case of Bose or Fermi fields the
Yang-Feldman formalism is equivalent to the standard Feynman-Dyson
diagram technique for $T$-product and one has the standard
 $T$-product expression for S-matrix in interaction representation
\cite{BD}.
In the Boltzmann field theory we have a more general field algebra than
ordinary Bose or Fermi  algebras and as a result one cannot
use the standard interaction picture. An appropriated interaction
representation  for half planar diagrams,
which form a subclass of planar diagrams,
has been constructed in \cite{AAV} (see also this volume).

The goal of this talk is to present
an interaction representation for the master field obtained in
\cite{AVM}. This new representation is more convenient
for  study of some aspects of the master field than the
Yang-Feldman formalism. In particular, this  representation
seems  more suitable
for investigations of renormalizability \footnote{Renormalizability in
Yang-Feldman formalism for Bose systems was investigated by Dyson
\cite{Dy} and Khallen \cite{Kal} } and gauge invariance
\footnote{Gauge invariance of the planar approximation for Yang-Mills
theory has been studded recently in \cite{Jan}}.
Moreover, we hope that the existence of closed set of equations for the lowest
Green's function for the simplest interaction representation for Boltzmann
theory permits to write a tractable non-perturbative
recursive set of non-linear integral-differential equations summing up planar
diagrams. This subject requires a further investigation.


Let us remain the main result of \cite{AVM}. There  it has been considered
 $U(N)$-invariant
Wightman's functions in the Yang-Feldman formalism,
\begin {equation} 
                                                          \label {1.4}
W(x_{1},...,x_{k})=
\frac{1}{N^{1+\frac{k}{2}}}<0|\tr(M(x_{1})...M(x_{k}))|0>
\end   {equation} 
where $M (x)=(M_{ij} (x)),$ $ i,j=1,...,N$ is an Hermitian
scalar matrix field  in the 4-dimensional
Minkowski space-time with the field equations
\begin {equation} 
                                                          \label {1.1e}
(\Box + m^{2})M (x)=J(x),~~
J(x)=-\sum \frac{c_{k}}{N^{-1+k/2}} M^{k-1} (x)
\end   {equation} 
where $c_{k}$ are the coupling constants which have no a dependence
on $N$.
One integrates eq (\ref {1.1e}) to get the Yang-Feldman equation
\cite {YF,BD}
\begin {equation} 
                                                          \label {1.3}
M(x)=M^{(in)}(x)+\int D^{ret}(x-y)J(y)dy
\end   {equation} 
where $D^{ret}(x)$ is the retarded Green function for the Klein-Gordon
equation,
$$
D^{ret}(x)=
\frac{1}{(2\pi)^{4}}\int \frac{e^{-ikx}}{m^{2}-k^{2}-
i\epsilon k^{0}}dk
$$
and  $M^{(in)}(x)$ is a  free Bose field
\begin {equation} 
                                                          \label {1.3'}
[M_{kl}^{(in)}(x),M_{l'k'}^{(in)}(y)]|_{x_{0}=y_{0}}=
\delta _{k,k'} \delta _{i,i'}D^{-}(x-y)
\end   {equation} 
$|0>$ is the Fock vacuum for the free field $M^{(in)}(x)$.

   In \cite{AVM} it has been show
that the limit of functions (\ref {1.4})  when $N\to \infty$ can be expressed
in terms of a quantum field $\phi (x)$ (the master field) which is a solution
of the equation
\begin {equation}
                                                     \label {1.5}
\phi(x)=\phi^{(in)}(x)+\int D^{ret}(x-y)j(y)dy,~~
j(x)=-\sum c_{k} \phi ^{k-1} (x)
\end   {equation} 
The  field  $\phi(x)$ does not have matrix indexes.
The free scalar Boltzmannian field $\phi^{(in)}(x)$
is given by
\begin {equation} 
                                                          \label {1.7}
\phi^{(in)}(x)=\frac{1}{(2\pi)^{3/2}}\int \frac{d^{3}k}
{\sqrt{2\omega (k)}}(a^*(k)e^{ikx}+a(k)e^{-ikx}) ,
\end   {equation} 
where $\omega (k)= \sqrt{k^{2}+m^{2}}$.  It satisfies to
the Klein-Gordon equation
$$(\Box + m^2)\phi^{(in)} (x)=0$$
and it is an operator in the Boltzmannian Fock space
with relations
\begin {equation} 
                                                          \label {1.8}
a(k)a^*(k')=\delta^{(3)}(k-k'),~~ \mbox {and}
 ~ ~ a(k)|\Omega_0)=0
\end   {equation} 

The relation between the large N invariant Wightman's functions and
the Boltzmannian Wightman functions is the following
\begin {equation} 
                                                          \label {1.1}
\lim _{N \to \infty}
\frac{1}{N^{1+\frac{k}{2}}}<0|\tr(M(x_1)...M(x_k))|0>
=(\Omega_{0}|\phi (x_{1})...\phi (x_{k})|\Omega_{0})
\end   {equation} 
where   the field $M (x)$ is defined by  (\ref {1.3})
and (\ref {1.3'})
and  $\phi (x)$ is defined by (\ref {1.5}),(\ref {1.7}) and  (\ref {1.8}).

In this talk I am going to show that for the master field
exists the following interaction representation
for connected Wightman's functions
\begin {equation} 
                                                          \label {1}
(\Omega_{0}|\phi (x_{1})...\phi (x_{k})|\Omega_{0})=
(0|\phi _{d}(x_{k})...\phi _{d}(x_{1})\frac{1}{1+\int d^{D}x{\cal
L}(\bar {\phi _{d}}(x),\phi _{d}(x),\phi _{u}(x))}|0)
\end   {equation}

Fields $\bar {\phi _{d}}(x)$, $\phi _{d}(x)$ and $\phi
_{u}(x)$ are Boltzmann fields.
The first two fields we call the down fields and $\phi
_{u}(x)$ field we call the up field. The meaning of this terminology will be
clear below.   The up and down fields are mutually commute and
they are sums of creation and annihilation parts
\begin {equation} 
                                                             \label {3}
\phi_{d}(x)=\phi ^{+}_{d}(x)+\phi ^{-}_{d}(x),
~{\bar \phi}_{d}(x)={\bar \phi} ^{+}_{d}(x)+{\bar \phi} ^{-}_{d}(x),
~
\phi _{u}(x)=\phi ^{+}_{u}(x)+\phi ^{-}_{u}(x),
\end   {equation} 
\begin {equation} 
                                                            \label {4'}
\phi ^{-}_{u}(x)|0)=\phi ^{-}_{d}(x)|0)=\bar {\phi ^{-}}_{u}(x)|0)=0
\end   {equation} 
\begin {equation} 
                                                            \label {4''}
(0|\phi ^{+}_{u}(x)=(0|\phi ^{+}_{d}(x)=(0|\bar {\phi ^{+}}_{u}(x)=0,
\end   {equation} 
and they satisfy the
following operator algebra
\begin {equation} 
                                                            \label {5}
\bar{ \phi }^{-}_{d}(x)\phi ^{+}_{d}(y)=D^{ret}(x-y),
~\phi ^{-}_{d}(y)\bar{\phi ^{+}}_{d}(x)=D^{ret}(x-y),
\end   {equation} 
\begin {equation} 
                                                         \label {6}
\phi ^{-}_{u}(x)\phi ^{+}_{u}(y)=D^{-}(y-x),~
\phi ^{-}_{d}(x)\phi ^{+}_{d}(y)=
{\bar\phi }^{-}_{d}(x){\bar \phi }^{+}_{d}(y)=0
\end   {equation} 
This algebra has a realization in the tensor product of ${\cal H}_{u}$
and ${\cal B}_{d}$ spaces. ${\cal H}_{u}$  is the Boltzmannian Fock space.
${\cal B}_{d}$  is a linear space with a  bilinear
form. This linear space is spanned on linear combinations
of products of creation part of down fields applying to the vacuum $|0)$

$$
(\phi ^{+}_{d}(y_{1}))^{k_{1}}
({\bar\phi }^{+}_{d}(z_{1})^{p_{1}}...
(\phi ^{+}_{d}(y_{l}))^{k_{l}}
({\bar\phi }^{+}_{d}(z_{l})^{p_{l}}|0).$$

Algebra (\ref{5}),(\ref{6}) is invariant under the following $*$
operation
\begin {equation} 
                                                         \label {7'}
(\phi ^{-}_{d}(x))^{*}=\phi ^{+}_{d}(y),~~
({\bar\phi }^{-}_{d}(x))^{*}={\bar \phi }^{+}_{d}(y),
~~(\phi ^{-}_{u}(x))^{*}=\phi ^{+}_{u}(y),
\end   {equation} 

Lagrangian  ${\cal L}$ in formula (\ref {1}) is defined by the form of the
 matrix  field Lagrangian. For example,
for the simplest case of the cubic interaction
\begin {equation} 
                                                          \label {8}
L= \frac{1}{2}\tr (\partial _{\mu} M)^{2}+
\frac{1}{2}m^{2}\tr (M)^{2}+ \frac{g}{3\cdot N^{1/2}}\tr M^{3},
\end   {equation} 
${\cal L}$ has a form
\begin {equation} 
                                                          \label {9}
{\cal L}=((\Box + m^{2})\bar{\phi }_{d}(x))\phi _{u}+
g[\bar{\phi }_{d}(x)(\phi ^{2}_{d}(x)+\phi _{d}(x)\phi _{u}(x)+
\phi ^{2}_{u}(x))+\phi _{u}(x)\phi _{d}(x)\bar{\phi }_{d}(x)]
\end   {equation} 
Note that ${\cal L}$ is not invariant under $*$-operation.
This is not surprising if one takes into account that the construction
of the master field in the 0-dimensional case also
deals with the non-Hermitial master field \cite{GG}.

Together with (\ref{1}) the representation (\ref{1.1})
gives

\begin {equation} 
                                                          \label {IR}
\lim _{N \to \infty}
\frac{1}{N^{1+\frac{k}{2}}}<0|\tr(M(x_1)...M(x_k))|0>
\end   {equation}
$$=(0|\phi _{d}(x_{k})...\phi _{d}(x_{1})\frac{1}{1+\int d^{D}x{\cal
L}(\bar {\phi _{d}}(x),\phi _{d}(x),\phi _{u}(x))}|0)$$

An analytical origin of the  representation (\ref{IR}) is the representation
(\ref{1.1}).
A topological origin of this representations is that any planar diagram
can be
redrawn so that all vertexes lie on some straight line dividing the plane on
which diagram is drown on two half planes, the upper and down half plane.
All lines in respect of this division of the plane are divided on
upper and down lines. Upper and down lines have no intersections.
There are many possibilities to perform a decomposition of the lines of  the
planar diagram into upper and down lines. We will prove that the special form
(\ref {1}) with relations  (\ref {3})-(\ref {6}) solve the 
corresponding combinatorial problem in the proper way.

 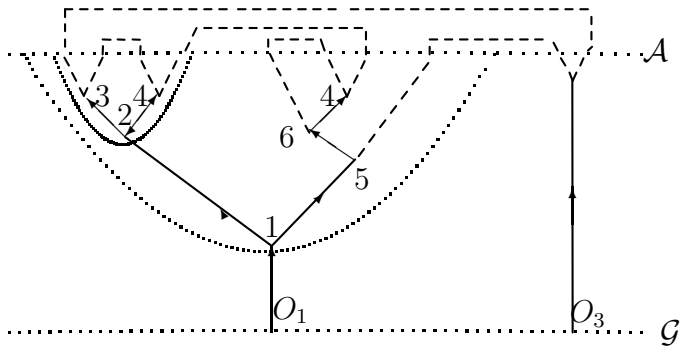
\begin{figure}
\begin{center}
\unitlength=0.50mm
\special{em:linewidth 0.8pt}
\linethickness{0.8pt}
\begin{picture}(176.00,107.00)
\put(70.00,21.00){\vector(0,1){23.00}}
\put(20.00,84.00){\line(-1,2){1.00}}
\put(20.00,84.00){\line(1,2){1.00}}
\put(40.00,84.00){\line(-1,2){1.00}}
\put(40.00,84.00){\line(1,2){1.00}}
\put(80.00,74.00){\line(-1,2){1.00}}
\bezier{70}(12.00,94.00)(30.00,48.00)(48.00,94.00)
\bezier{80}(7.00,89.00)(70.00,-4.00)(125.00,89.00)
\bezier{4}(7.00,89.00)(5.00,93.00)(4.00,95.00)
\put(70.00,49.00){\makebox(0,0)[cc]{$1$}}
\put(75.00,27.00){\makebox(0,0)[cc]{$O_{1}$}}
\put(94.00,62.00){\makebox(0,0)[cc]{$5$}}
\put(74.00,73.00){\makebox(0,0)[cc]{$6$}}
\put(31.00,79.00){\makebox(0,0)[cc]{$2$}}
\put(25.00,84.00){\makebox(0,0)[cc]{$3$}}
\put(35.00,84.00){\makebox(0,0)[cc]{$4$}}
\bezier{2}(125.00,89.00)(126.00,90.00)(129.00,95.00)
\put(154.00,26.00){\makebox(0,0)[cc]{$O_{3}$}}
\put(20.30,84.15){\line(-1,2){0.95}}
\put(20.30,83.91){\line(1,3){0.72}}
\put(39.86,83.91){\line(-1,2){0.95}}
\put(37.95,87.72){\line(-1,2){0.95}}
\put(36.04,91.54){\line(-1,2){0.95}}
\put(40.10,83.91){\line(1,2){0.95}}
\put(42.01,87.72){\line(1,2){0.95}}
\put(43.91,91.54){\line(1,2){0.95}}
\put(89.85,84.00){\line(-1,2){1.00}}
\put(89.85,84.00){\line(1,2){1.00}}
\put(84.85,84.00){\makebox(0,0)[cc]{$4$}}
\put(89.71,83.91){\line(-1,2){0.95}}
\put(89.94,83.91){\line(1,2){0.95}}
\emline{40.10}{83.91}{1}{39.15}{85.81}{2}
\emline{38.19}{87.48}{3}{37.24}{89.39}{4}
\emline{36.28}{91.30}{5}{35.33}{93.21}{6}
\emline{40.10}{83.91}{7}{41.05}{85.81}{8}
\emline{41.77}{87.48}{9}{42.72}{89.39}{10}
\emline{43.68}{91.30}{11}{44.63}{93.21}{12}
\emline{79.93}{74.13}{13}{78.97}{76.04}{14}
\emline{78.02}{77.94}{15}{77.07}{79.85}{16}
\emline{76.11}{81.76}{17}{75.16}{83.67}{18}
\emline{74.20}{85.58}{19}{73.25}{87.48}{20}
\emline{72.30}{89.39}{21}{71.34}{91.30}{22}
\emline{70.39}{93.21}{23}{69.43}{95.12}{24}
\emline{89.94}{83.91}{25}{88.99}{85.81}{26}
\emline{88.04}{87.72}{27}{87.08}{89.63}{28}
\emline{86.13}{91.54}{29}{85.17}{93.45}{30}
\emline{89.94}{83.67}{31}{90.90}{85.81}{32}
\emline{91.85}{87.48}{33}{92.81}{89.39}{34}
\emline{93.76}{91.30}{35}{94.71}{93.21}{36}
\emline{93.04}{67.93}{37}{94.48}{69.84}{38}
\emline{95.91}{71.74}{39}{97.34}{73.65}{40}
\emline{98.77}{75.56}{41}{100.20}{77.47}{42}
\emline{101.63}{79.38}{43}{103.06}{81.28}{44}
\emline{104.49}{83.19}{45}{105.92}{85.10}{46}
\emline{107.35}{87.01}{47}{108.79}{88.91}{48}
\emline{110.22}{90.82}{49}{111.65}{92.73}{50}
\emline{20.00}{84.00}{51}{18.90}{86.02}{52}
\emline{18.00}{88.00}{53}{16.88}{90.07}{54}
\emline{16.00}{92.00}{55}{14.85}{94.12}{56}
\emline{25.00}{95.00}{57}{25.10}{97.15}{58}
\emline{25.00}{99.00}{59}{27.13}{98.84}{60}
\emline{29.00}{99.00}{61}{30.84}{98.84}{62}
\emline{33.00}{99.00}{63}{34.89}{98.84}{64}
\emline{20.00}{84.00}{65}{20.93}{86.02}{66}
\emline{22.00}{88.00}{67}{22.95}{90.07}{68}
\emline{24.00}{92.00}{69}{24.98}{94.12}{70}
\emline{35.00}{95.00}{71}{35.10}{97.15}{72}
\emline{69.00}{97.00}{73}{69.00}{99.00}{74}
\emline{71.00}{99.00}{75}{73.00}{99.00}{76}
\emline{75.00}{99.00}{77}{77.00}{99.00}{78}
\emline{79.00}{99.00}{79}{81.00}{99.00}{80}
\emline{83.00}{99.00}{81}{84.00}{96.00}{82}
\emline{47.00}{96.00}{83}{48.00}{98.00}{84}
\emline{49.00}{100.00}{85}{50.00}{102.00}{86}
\emline{52.00}{102.00}{87}{54.00}{102.00}{88}
\emline{56.00}{102.00}{89}{58.00}{102.00}{90}
\emline{60.00}{102.00}{91}{62.00}{102.00}{92}
\emline{64.00}{102.00}{93}{66.00}{102.00}{94}
\emline{68.00}{102.00}{95}{70.00}{102.00}{96}
\emline{72.00}{102.00}{97}{74.00}{102.00}{98}
\emline{76.00}{102.00}{99}{78.00}{102.00}{100}
\emline{80.00}{102.00}{101}{82.00}{102.00}{102}
\emline{84.00}{102.00}{103}{86.00}{102.00}{104}
\emline{88.00}{102.00}{105}{90.00}{102.00}{106}
\emline{92.00}{102.00}{107}{94.00}{102.00}{108}
\emline{95.00}{101.00}{109}{95.00}{99.00}{110}
\emline{95.00}{97.00}{111}{95.00}{95.00}{112}
\emline{112.00}{95.00}{113}{112.00}{97.00}{114}
\emline{112.00}{99.00}{115}{114.00}{99.00}{116}
\emline{116.00}{99.00}{117}{118.00}{99.00}{118}
\emline{120.00}{99.00}{119}{122.00}{99.00}{120}
\emline{124.00}{99.00}{121}{126.00}{99.00}{122}
\emline{128.00}{99.00}{123}{130.00}{99.00}{124}
\emline{132.00}{99.00}{125}{134.00}{99.00}{126}
\emline{136.00}{99.00}{127}{138.00}{99.00}{128}
\emline{140.00}{99.00}{129}{142.00}{99.00}{130}
\emline{155.00}{99.00}{131}{155.00}{101.00}{132}
\emline{155.00}{103.00}{133}{155.00}{105.00}{134}
\emline{154.00}{107.00}{135}{152.00}{107.00}{136}
\emline{124.00}{107.00}{137}{126.00}{107.00}{138}
\emline{128.00}{107.00}{139}{130.00}{107.00}{140}
\emline{132.00}{107.00}{141}{134.00}{107.00}{142}
\emline{136.00}{107.00}{143}{138.00}{107.00}{144}
\emline{140.00}{107.00}{145}{142.00}{107.00}{146}
\emline{144.00}{107.00}{147}{146.00}{107.00}{148}
\emline{148.00}{107.00}{149}{150.00}{107.00}{150}
\emline{121.00}{107.00}{151}{119.00}{107.00}{152}
\emline{91.00}{107.00}{153}{93.00}{107.00}{154}
\emline{95.00}{107.00}{155}{97.00}{107.00}{156}
\emline{99.00}{107.00}{157}{101.00}{107.00}{158}
\emline{103.00}{107.00}{159}{105.00}{107.00}{160}
\emline{107.00}{107.00}{161}{109.00}{107.00}{162}
\emline{111.00}{107.00}{163}{113.00}{107.00}{164}
\emline{115.00}{107.00}{165}{117.00}{107.00}{166}
\emline{61.00}{107.00}{167}{63.00}{107.00}{168}
\emline{65.00}{107.00}{169}{67.00}{107.00}{170}
\emline{69.00}{107.00}{171}{71.00}{107.00}{172}
\emline{73.00}{107.00}{173}{75.00}{107.00}{174}
\emline{77.00}{107.00}{175}{79.00}{107.00}{176}
\emline{81.00}{107.00}{177}{83.00}{107.00}{178}
\emline{85.00}{107.00}{179}{87.00}{107.00}{180}
\emline{58.00}{107.00}{181}{56.00}{107.00}{182}
\emline{28.00}{107.00}{183}{30.00}{107.00}{184}
\emline{32.00}{107.00}{185}{34.00}{107.00}{186}
\emline{36.00}{107.00}{187}{38.00}{107.00}{188}
\emline{40.00}{107.00}{189}{42.00}{107.00}{190}
\emline{44.00}{107.00}{191}{46.00}{107.00}{192}
\emline{48.00}{107.00}{193}{50.00}{107.00}{194}
\emline{52.00}{107.00}{195}{54.00}{107.00}{196}
\emline{15.00}{107.00}{197}{17.00}{107.00}{198}
\emline{19.00}{107.00}{199}{21.00}{107.00}{200}
\emline{23.00}{107.00}{201}{25.00}{107.00}{202}
\emline{15.00}{107.00}{203}{15.00}{104.92}{204}
\emline{15.00}{103.00}{205}{15.00}{101.17}{206}
\emline{15.00}{99.00}{207}{15.00}{97.00}{208}
\bezier{60}(0.00,21.00)(80.00,22.00)(168.00,21.00)
\put(172.00,96.00){\makebox(0,0)[cc]{${\cal A}$}}
\put(176.00,21.00){\makebox(0,0)[cc]{${\cal G}$}}
\bezier{50}(0.00,95.00)(80.00,96.00)(168.00,95.00)
\emline{70.00}{44.00}{209}{92.18}{67.03}{210}
\put(80.00,54.00){\vector(1,1){3.84}}
\put(92.00,67.00){\vector(-3,2){11.96}}
\put(80.04,74.24){\vector(1,1){9.86}}
\put(31.11,73.10){\vector(-1,-1){0.00}}
\put(31.11,73.10){\vector(3,4){8.35}}
\put(31.00,73.00){\vector(-1,1){10.14}}
\emline{150.00}{21.00}{211}{149.85}{87.90}{212}
\emline{150.00}{88.00}{213}{150.98}{90.17}{214}
\emline{152.00}{92.00}{215}{152.88}{93.97}{216}
\emline{154.00}{96.00}{217}{155.16}{97.00}{218}
\emline{150.00}{88.00}{219}{149.09}{90.17}{220}
\emline{148.00}{92.00}{221}{146.81}{93.97}{222}
\emline{146.00}{96.00}{223}{144.91}{98.14}{224}
\put(150.00,49.67){\vector(0,1){10.00}}
\emline{70.00}{44.00}{225}{31.00}{73.00}{226}
\put(56.67,53.33){\vector(-1,2){0.33}}
\end{picture}

\end{center}
\vspace{-5mm}
\caption{Diagram representing two-point Wightman's function}\label{F1}
\end{figure}

\begin{figure}
\begin{center}
\unitlength=1.50mm
\special{em:linewidth 0.4pt}
\linethickness{0.4pt}
\begin{picture}(42.94,16.14)
\put(10.00,5.00){\line(0,1){5.00}}
\put(10.00,10.00){\line(1,2){3.07}}
\put(13.07,16.00){\line(0,0){0.00}}
\put(10.00,10.00){\line(-1,2){3.07}}
\put(10.00,10.00){\vector(-1,2){2.00}}
\put(10.00,10.00){\vector(1,2){2.00}}
\put(10.00,5.00){\vector(0,1){4.07}}
\put(20.00,5.00){\line(0,1){5.00}}
\put(20.00,10.00){\line(-1,2){3.07}}
\put(20.00,10.00){\vector(-1,2){2.00}}
\put(20.00,5.00){\vector(0,1){4.07}}
\put(20.00,9.07){\vector(0,0){0.00}}
\put(30.00,5.00){\vector(0,1){4.07}}
\put(40.00,5.00){\vector(0,1){4.07}}
\emline{20.00}{9.00}{1}{21.07}{11.07}{2}
\emline{22.00}{13.00}{3}{23.07}{15.07}{4}
\emline{30.00}{9.00}{5}{29.07}{11.07}{6}
\emline{28.00}{13.00}{7}{27.07}{15.07}{8}
\put(40.00,9.00){\line(-1,2){0.91}}
\put(38.00,13.00){\line(-1,2){0.96}}
\put(40.00,9.00){\line(1,2){1.02}}
\put(42.00,13.00){\line(1,2){0.94}}
\emline{40.00}{9.00}{9}{38.96}{10.97}{10}
\emline{38.00}{13.00}{11}{37.04}{14.95}{12}
\emline{40.00}{9.00}{13}{41.02}{10.97}{14}
\emline{42.00}{13.00}{15}{42.94}{14.95}{16}
\emline{30.00}{9.00}{17}{33.07}{16.00}{18}
\put(31.47,12.40){\vector(1,3){0.53}}
\end{picture}

\end{center}
\vspace{-5mm}
\caption{Types of vertexes in diagrams representing  Wightman's
functions in the Yang-Feldman formalism for the cubic
interaction }\label{F2} \end{figure}
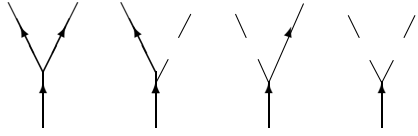

\begin{figure}
\begin{center}
\unitlength=0.50mm
\special{em:linewidth 0.8pt}
\linethickness{0.8pt}
\begin{picture}(200.00,215.14)
\put(94.00,138.00){\vector(0,1){23.00}}
\put(44.00,201.00){\line(-1,2){1.00}}
\put(44.00,201.00){\line(1,2){1.00}}
\put(64.00,201.00){\line(-1,2){1.00}}
\put(64.00,201.00){\line(1,2){1.00}}
\put(104.00,191.00){\line(-1,2){1.00}}
\bezier{70}(36.00,211.00)(54.00,165.00)(72.00,211.00)
\bezier{80}(31.00,206.00)(94.00,113.00)(149.00,206.00)
\bezier{4}(31.00,206.00)(29.00,210.00)(28.00,212.00)
\special{em:linewidth 0.4pt}
\linethickness{0.4pt}
\put(41.00,36.00){\oval(10.00,10.00)[b]}
\put(94.50,36.00){\oval(97.00,40.00)[b]}
\put(129.50,36.00){\oval(27.00,18.00)[b]}
\put(121.00,36.00){\oval(10.00,10.00)[b]}
\put(133.00,27.00){\vector(-1,0){5.00}}
\put(86.00,16.00){\vector(1,0){17.00}}
\special{em:linewidth 0.8pt}
\linethickness{0.8pt}
\put(65.00,36.00){\line(-1,2){1.00}}
\put(65.00,36.00){\line(1,2){1.00}}
\put(84.00,36.00){\line(-1,2){1.00}}
\put(84.00,36.00){\line(1,2){1.00}}
\put(126.00,36.00){\line(-1,2){1.00}}
\put(126.00,36.00){\line(1,2){1.00}}
\put(143.00,36.00){\line(0,1){2.00}}
\put(143.00,40.00){\line(0,1){2.00}}
\put(143.00,44.00){\line(0,1){2.00}}
\put(143.00,36.00){\line(0,1){2.00}}
\put(143.00,40.00){\line(0,1){2.00}}
\put(143.00,44.00){\line(0,1){2.00}}
\put(116.00,36.00){\line(0,1){2.00}}
\put(116.00,40.00){\line(0,1){2.00}}
\put(116.00,44.00){\line(0,1){2.00}}
\put(116.00,36.00){\line(0,1){2.00}}
\put(116.00,40.00){\line(0,1){2.00}}
\put(116.00,44.00){\line(0,1){2.00}}
\put(49.00,39.00){\makebox(0,0)[cc]{$^{1^{\prime}}$}}
\put(71.00,39.00){\makebox(0,0)[cc]{$^{3^{\prime}}$}}
\put(58.00,39.00){\makebox(0,0)[cc]{$^{2^{\prime}}$}}
\put(90.00,39.00){\makebox(0,0)[cc]{$^{4^{\prime}}$}}
\put(113.00,39.00){\makebox(0,0)[cc]{$^{6^{\prime}}$}}
\put(131.00,39.00){\makebox(0,0)[cc]{$~^{7^{\prime}}$}}
\put(146.00,39.00){\makebox(0,0)[cc]{$~~^{5^{\prime}}$}}
\put(37.00,39.00){\makebox(0,0)[cc]{$^{O^{\prime}_{1}}$}}
\put(94.00,166.00){\makebox(0,0)[cc]{$1$}}
\put(99.00,144.00){\makebox(0,0)[cc]{$~O_{1}$}}
\put(118.00,179.00){\makebox(0,0)[cc]{$5$}}
\put(98.00,190.00){\makebox(0,0)[cc]{$6$}}
\put(55.00,196.00){\makebox(0,0)[cc]{$2$}}
\put(49.00,201.00){\makebox(0,0)[cc]{$3$}}
\put(59.00,201.00){\makebox(0,0)[cc]{$4$}}
\emline{65.00}{36.00}{1}{63.99}{38.08}{2}
\emline{63.99}{38.08}{3}{63.99}{38.08}{4}
\emline{63.00}{40.00}{5}{62.06}{41.95}{6}
\emline{61.00}{44.00}{7}{60.13}{46.09}{8}
\emline{65.00}{36.00}{9}{65.93}{38.08}{10}
\emline{67.00}{40.00}{11}{68.14}{41.95}{12}
\emline{69.00}{44.00}{13}{70.07}{46.09}{14}
\emline{84.00}{36.00}{15}{83.05}{38.08}{16}
\emline{82.00}{40.00}{17}{81.12}{41.95}{18}
\emline{80.00}{44.00}{19}{78.91}{46.09}{20}
\emline{84.00}{36.00}{21}{84.98}{38.08}{22}
\emline{86.00}{40.00}{23}{86.91}{41.95}{24}
\emline{88.00}{44.00}{25}{89.12}{46.09}{26}
\emline{126.00}{36.00}{27}{125.02}{38.08}{28}
\emline{124.00}{40.00}{29}{123.09}{41.95}{30}
\emline{122.00}{44.00}{31}{120.88}{46.09}{32}
\emline{126.00}{36.00}{33}{126.96}{38.08}{34}
\emline{128.00}{40.00}{35}{128.89}{41.95}{36}
\emline{130.00}{44.00}{37}{131.10}{46.09}{38}
\bezier{2}(149.00,206.00)(150.00,207.00)(153.00,212.00)
\special{em:linewidth 0.4pt}
\linethickness{0.4pt}
\put(100.50,35.50){\oval(145.00,61.00)[b]}
\special{em:linewidth 0.8pt}
\linethickness{0.8pt}
\put(178.00,143.00){\makebox(0,0)[cc]{$O_{3}$}}
\emline{173.00}{36.00}{39}{172.00}{38.00}{40}
\emline{171.00}{40.00}{41}{170.00}{42.00}{42}
\emline{169.00}{44.00}{43}{168.00}{46.00}{44}
\emline{173.00}{36.00}{45}{174.00}{38.00}{46}
\emline{175.00}{40.00}{47}{176.00}{42.00}{48}
\emline{177.00}{44.00}{49}{178.00}{46.00}{50}
\put(82.00,5.00){\vector(1,0){21.00}}
\put(26.00,39.00){\makebox(0,0)[cc]{$^{O^{\prime}_{3}}$}}
\put(44.30,201.15){\line(-1,2){0.95}}
\put(44.30,200.91){\line(1,3){0.72}}
\put(63.86,200.91){\line(-1,2){0.95}}
\put(61.95,204.72){\line(-1,2){0.95}}
\put(60.04,208.54){\line(-1,2){0.95}}
\put(64.10,200.91){\line(1,2){0.95}}
\put(66.01,204.72){\line(1,2){0.95}}
\put(67.91,208.54){\line(1,2){0.95}}
\put(113.85,201.00){\line(-1,2){1.00}}
\put(113.85,201.00){\line(1,2){1.00}}
\put(108.85,201.00){\makebox(0,0)[cc]{$4$}}
\put(113.71,200.91){\line(-1,2){0.95}}
\put(113.94,200.91){\line(1,2){0.95}}
\emline{64.10}{200.91}{51}{63.15}{202.81}{52}
\emline{62.19}{204.48}{53}{61.24}{206.39}{54}
\emline{60.28}{208.30}{55}{59.33}{210.21}{56}
\emline{64.10}{200.91}{57}{65.05}{202.81}{58}
\emline{65.77}{204.48}{59}{66.72}{206.39}{60}
\emline{67.68}{208.30}{61}{68.63}{210.21}{62}
\emline{103.93}{191.13}{63}{102.97}{193.04}{64}
\emline{102.02}{194.94}{65}{101.07}{196.85}{66}
\emline{100.11}{198.76}{67}{99.16}{200.67}{68}
\emline{98.20}{202.58}{69}{97.25}{204.48}{70}
\emline{96.30}{206.39}{71}{95.34}{208.30}{72}
\emline{94.39}{210.21}{73}{93.43}{212.12}{74}
\emline{113.94}{200.91}{75}{112.99}{202.81}{76}
\emline{112.04}{204.72}{77}{111.08}{206.63}{78}
\emline{110.13}{208.54}{79}{109.17}{210.45}{80}
\emline{113.94}{200.67}{81}{114.90}{202.81}{82}
\emline{115.85}{204.48}{83}{116.81}{206.39}{84}
\emline{117.76}{208.30}{85}{118.71}{210.21}{86}
\emline{117.04}{184.93}{87}{118.48}{186.84}{88}
\emline{119.91}{188.74}{89}{121.34}{190.65}{90}
\emline{122.77}{192.56}{91}{124.20}{194.47}{92}
\emline{125.63}{196.38}{93}{127.06}{198.28}{94}
\emline{128.49}{200.19}{95}{129.92}{202.10}{96}
\emline{131.35}{204.01}{97}{132.79}{205.91}{98}
\emline{134.22}{207.82}{99}{135.65}{209.73}{100}
\emline{44.00}{201.00}{101}{42.90}{203.02}{102}
\emline{42.00}{205.00}{103}{40.88}{207.07}{104}
\emline{40.00}{209.00}{105}{38.85}{211.12}{106}
\emline{44.00}{201.00}{107}{44.93}{203.02}{108}
\emline{46.00}{205.00}{109}{46.95}{207.07}{110}
\emline{48.00}{209.00}{111}{48.98}{211.12}{112}
\bezier{60}(24.00,138.00)(104.00,139.00)(192.00,138.00)
\put(196.00,213.00){\makebox(0,0)[cc]{${\cal A}$}}
\put(200.00,138.00){\makebox(0,0)[cc]{${\cal G}$}}
\special{em:linewidth 0.4pt}
\linethickness{0.4pt}
\put(51.50,36.00){\oval(11.00,10.00)[b]}
\put(70.50,36.00){\oval(27.00,26.00)[b]}
\put(40.00,31.00){\vector(1,0){1.92}}
\put(49.00,31.00){\vector(1,0){2.92}}
\put(67.00,23.00){\vector(1,0){4.92}}
\put(129.50,103.00){\oval(27.00,18.00)[b]}
\put(121.00,103.00){\oval(10.00,10.00)[b]}
\special{em:linewidth 0.8pt}
\linethickness{0.8pt}
\bezier{100}(24.00,36.00)(107.00,35.00)(189.00,36.00)
\put(65.00,103.00){\line(-1,2){1.00}}
\put(65.00,103.00){\line(1,2){1.00}}
\put(84.00,103.00){\line(-1,2){1.00}}
\put(84.00,103.00){\line(1,2){1.00}}
\put(126.00,103.00){\line(-1,2){1.00}}
\put(126.00,103.00){\line(1,2){1.00}}
\put(143.00,103.00){\line(0,1){2.00}}
\put(143.00,107.00){\line(0,1){2.00}}
\put(143.00,111.00){\line(0,1){2.00}}
\put(143.00,103.00){\line(0,1){2.00}}
\put(143.00,107.00){\line(0,1){2.00}}
\put(143.00,111.00){\line(0,1){2.00}}
\put(116.00,103.00){\line(0,1){2.00}}
\put(116.00,107.00){\line(0,1){2.00}}
\put(116.00,111.00){\line(0,1){2.00}}
\put(116.00,103.00){\line(0,1){2.00}}
\put(116.00,107.00){\line(0,1){2.00}}
\put(116.00,111.00){\line(0,1){2.00}}
\put(133.00,94.00){\vector(-1,0){5.00}}
\put(71.00,106.00){\makebox(0,0)[cc]{$^{3^{\prime}}$}}
\put(90.00,106.00){\makebox(0,0)[cc]{$^{4^{\prime}}$}}
\put(113.00,106.00){\makebox(0,0)[cc]{$^{6^{\prime}}$}}
\put(131.00,106.00){\makebox(0,0)[cc]{$^{7^{\prime}}$}}
\put(146.00,106.00){\makebox(0,0)[cc]{$^{5^{\prime}}$}}
\emline{65.00}{103.00}{113}{63.99}{105.08}{114}
\emline{63.99}{105.08}{115}{63.99}{105.08}{116}
\emline{63.00}{107.00}{117}{62.06}{108.95}{118}
\emline{61.00}{111.00}{119}{60.13}{113.09}{120}
\emline{65.00}{103.00}{121}{65.93}{105.08}{122}
\emline{67.00}{107.00}{123}{68.14}{108.95}{124}
\emline{69.00}{111.00}{125}{70.07}{113.09}{126}
\emline{84.00}{103.00}{127}{83.05}{105.08}{128}
\emline{82.00}{107.00}{129}{81.12}{108.95}{130}
\emline{80.00}{111.00}{131}{78.91}{113.09}{132}
\emline{84.00}{103.00}{133}{84.98}{105.08}{134}
\emline{86.00}{107.00}{135}{86.91}{108.95}{136}
\emline{88.00}{111.00}{137}{89.12}{113.09}{138}
\emline{126.00}{103.00}{139}{125.02}{105.08}{140}
\emline{124.00}{107.00}{141}{123.09}{108.95}{142}
\emline{122.00}{111.00}{143}{120.88}{113.09}{144}
\emline{126.00}{103.00}{145}{126.96}{105.08}{146}
\emline{128.00}{107.00}{147}{128.89}{108.95}{148}
\emline{130.00}{111.00}{149}{131.10}{113.09}{150}
\emline{173.00}{103.00}{151}{172.00}{105.00}{152}
\emline{171.00}{107.00}{153}{170.00}{109.00}{154}
\emline{169.00}{111.00}{155}{168.00}{113.00}{156}
\emline{173.00}{103.00}{157}{174.00}{105.00}{158}
\emline{175.00}{107.00}{159}{176.00}{109.00}{160}
\emline{177.00}{111.00}{161}{178.00}{113.00}{162}
\bezier{60}(24.00,103.00)(107.00,102.00)(189.00,103.00)
\put(74.00,93.00){\vector(-1,1){9.17}}
\put(74.00,93.00){\vector(1,1){9.17}}
\special{em:linewidth 0.4pt}
\linethickness{0.4pt}
\put(109.00,92.50){\oval(70.00,17.00)[lb]}
\put(105.50,93.00){\oval(75.00,18.00)[rb]}
\emline{143.00}{92.00}{163}{143.17}{102.08}{164}
\put(61.00,36.00){\oval(8.00,10.00)[b]}
\put(60.00,31.00){\vector(1,0){2.00}}
\put(121.00,84.00){\vector(1,0){5.92}}
\put(98.00,84.00){\vector(-1,0){6.08}}
\special{em:linewidth 0.8pt}
\linethickness{0.8pt}
\bezier{100}(24.00,66.00)(107.00,65.00)(189.00,66.00)
\bezier{50}(60.00,102.00)(74.00,81.00)(89.00,102.00)
\bezier{116}(43.00,103.00)(108.00,36.00)(162.00,103.00)
\put(108.00,88.00){\makebox(0,0)[cc]{$1$}}
\put(74.00,96.00){\makebox(0,0)[cc]{$^{2}$}}
\put(44.00,106.00){\makebox(0,0)[cc]{$g_{1}$}}
\put(59.00,106.00){\makebox(0,0)[cc]{$g_{2}$}}
\bezier{50}(24.00,212.00)(104.00,213.00)(192.00,212.00)
\put(121.00,31.00){\vector(1,0){0.92}}
\emline{94.00}{161.00}{165}{116.18}{184.03}{166}
\put(104.00,171.00){\vector(1,1){3.84}}
\put(116.00,184.00){\vector(-3,2){11.96}}
\put(104.04,191.24){\vector(1,1){9.86}}
\put(55.11,190.10){\vector(-1,-1){0.00}}
\put(55.11,190.10){\vector(3,4){8.35}}
\put(55.00,190.00){\vector(-1,1){10.14}}
\emline{174.00}{138.00}{167}{173.85}{204.90}{168}
\emline{174.00}{205.00}{169}{174.98}{207.17}{170}
\emline{176.00}{209.00}{171}{176.88}{210.97}{172}
\emline{178.00}{213.00}{173}{179.16}{214.00}{174}
\emline{174.00}{205.00}{175}{173.09}{207.17}{176}
\emline{172.00}{209.00}{177}{170.81}{210.97}{178}
\emline{170.00}{213.00}{179}{168.91}{215.14}{180}
\emline{108.00}{65.00}{181}{108.03}{83.96}{182}
\put(108.00,74.00){\vector(0,1){0.03}}
\emline{173.00}{65.00}{183}{172.97}{103.04}{184}
\put(173.00,80.00){\vector(0,1){0.03}}
\put(174.00,166.67){\vector(0,1){10.00}}
\emline{94.00}{161.00}{185}{55.00}{190.00}{186}
\put(80.67,170.33){\vector(-1,2){0.33}}
\put(199.00,105.00){\makebox(0,0)[cc]{${\cal A}$}}
\put(199.00,66.00){\makebox(0,0)[cc]{${\cal G}$}}
\put(199.00,36.00){\makebox(0,0)[cc]{${\cal A}$}}
\put(120.29,97.99){\vector(1,0){1.04}}
\put(14.00,36.00){\makebox(0,0)[cc]{${\cal G}$}}
\end{picture}

\end{center}
\vspace{-5mm}
\caption{Trees contributing to two-point Wightman's function}\label{F3}
\end{figure}
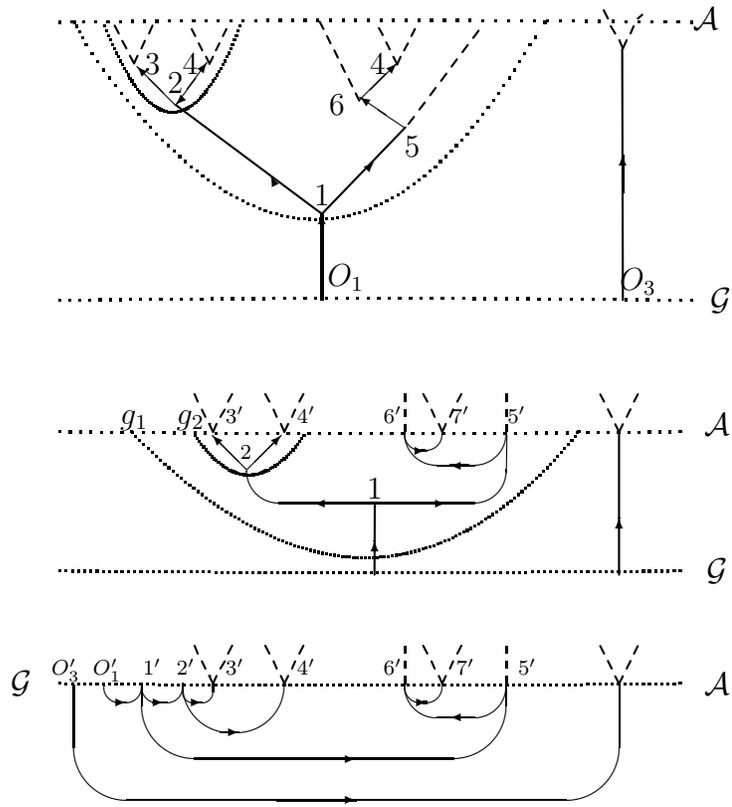

To this end let us fist make few comments about  an interpretation
of the RHS
of relation (\ref {1.1}) on diagrams. For simplicity we consider the
case of the cubic interaction. A perturbation expansion on the coupling
constant $c_{3}\equiv g$   of the RHS of (\ref
{1.1})      on  diagrams language
 can be represented     in the following way (see Fig.\ref{F1}).  A diagram
 representing a k-point Wightman's function
contains $k$  trees "planted" on the same "ground" and the trees have no
overlapping.
 On Fig.\ref{F1} ${\cal G}$-line represents the "ground" line
and there are two trees "planted" on the line ${\cal G}$.
Each tree represents a term of the perturbative solution of
operator equation (\ref{1.5}). Trees contain vertexes and links.
There are two type of links. We
call them branches and leaves, respectively. A branch  relates two
vertexes and a leaf is attached only to one vertex.
A branch of the tree represents
$D^{ret}(y_{i,j}-y_{i+1,j})$, where $y_{i,j}$  and $y_{i+1,j}$ are
coordinates of $i$-vertex and $i+1$-vertex belong $j$-tree, respectively.
We draw branches by solid lines.  Branches have direction. Each vertex of
the tree contains exactly one branch coming in and
several branches coming
out.  Trees contain also several leaves which are attached  to some
vertexes. Leaves represent $in$-fields. We draw leaves by  dash lines. So
there are several types of vertexes in the tree representing
the perturbative  solution of (\ref{1.5}) for the given interaction.
For the cubic interaction
there are four types of vertexes (see Fig.\ref{F2}).
Since equation (\ref{1.5}) is operator valued equation
and operators $\phi ^{in}$ do not commute it is essential to preserve
in the perturbation solution  the operator order
of  $\phi ^{in}$.
To keep the operator ordering we draw the tree so that its branches have no
overlapping.  All $in$-fields which are ordered according formula (\ref{1.5})
are also ordered in the diagram.  To make the ordering more obvious one can
deform the tree so that all vertex containing leaves are drown on some
auxiliary line (line ${\cal A}$ on Fig.\ref{F1}
and Fig.\ref{F3}a,b).

Note that the same picture is true also for diagrams representing the
 solution
of the Yang-Feldman equation for the usual Bose field.
The difference comes out then we consider expectation
values of products of several fields being solutions of the
Yang-Feldman equation. For example, the diagram Fig.\ref{F3}a also
describes the operator product of two Bose scalar fields being solutions
of the Yang-Feldman equation.

Since we consider expectation values in respect of the
$in$-vacuum all leaves
should be contracted and since we are doing calculations
in the Boltzmannian case the
contractions of leaves should  have no intersections with themselves.
By our construction leaves also have no intersections with
branches.  So, leaves are contracted in pairs 
without intersections with others
leaves or branches. We draw  contractions of leaves
by dash lines.
For the Bose case all contractions of leaves are
admissible. For example, the diagram Fig.\ref{F1} describes
the two-point Wigtman's function of the scalar field theory with the
cubic interaction, but together with this diagram
diagrams with overlapping contractions of leaves
also contribute.

As we just have mentioned we can deform the given diagram so that it
exists an
auxiliary line which crosses all leaves
(line ${\cal A}$ on Fig.\ref{F1}).  After
that we can redraw the diagram so that all leaves are started from this
auxiliary line. In this new picture (Fig.\ref{F3}b) all vertexes
containing at
least one leaf are on the auxiliary line ${\cal A}$.  To find a  point on
the auxiliary line corresponding to a vertex with only one leaf it is
enough to take a intersection of the corresponding dash line with the auxiliary
line. If a vertex contains two leaves one can take as a image of
the corresponding vertex the intersection of the auxiliary line with the left
leaf.  We note that after such redrawn contractions of the leaves are
represented as a set of  top "arcs" based on the auxiliary line.

Moreover, we can redraw each of diagrams representing (\ref{1.1})
 so that also all other
vertexes as well as the beginning of the external lines are drown on the
same auxiliary line ${\cal A}$ (see Fig.\ref{F3}c).
To make this fact obvious let us note that any tree ${\cal T}_{0,j}$
is dual to a set of bottom arcs \cite{KNN} so that to each 
part ${\cal T}_{i,j}$  of
tree ${\cal T}_{0,j}$
 corresponds an arc which surrounds  this tree ${\cal
T}_{i,j}$,
(${\cal T}_{i,j}$ is itself a tree; $i$ specified the $i$-vertex of
the $j$-tree,  see Fig.\ref{F3}a).

Moving each vertex ${\cal V}_{i,j}$ to some point belong a left
segment of the auxiliary line ${\cal A}$ which is cut out by the
arcs dual to the nearest trees we see
that all vertexes  of diagram are now on the auxiliary line
${\cal A}$.  For example, the image of vertex 1 on \ref{F3}b
must be on the segment $(g_1 g_2)$. Points $1^{\prime}$ and
$2^{\prime}$ on Fig.\ref{F3}c are images of vertexes 1 and 2.
Images of "ground" points (points $O'_{1}$ and $O'_{3}$
on Fig.\ref{F3}c) are also on the auxiliary line.
All branches of the given tree
are drawn now as solid
bottom arcs and these  arcs are started
from the images  of  "ground" points or vertexes
on the auxiliary line.
According this construction the solid arcs have no overlapping.
Some vertex are also
bases for dash arcs which
represent contractions of the leaves, or $in$-fields.
Note that after this redraw all
external lines are also started from the point lying on
the auxiliary
line ${\cal A}$ and they are arranged  in respect of the original ordering in the
ground line in the opposite way.
Since a solid
line  comes
in each vertex  and all vertex are situated
in the right of external points we can reconstruct a
part of the diagram that is  represented by dash lines as the result
of the calculation of the following
expectation value in the Boltzmann theory
\begin {equation} 
\label {1.9}
(0|\phi _{d}(x_{1})...\phi _{d}(x_{l})... {\bar \phi _{d}}(y_{i})
(\phi_{d}(y_{i}))^{k_{i}}...  \phi
_{d}(y_{l}){\bar \phi _{d}}(y_{l})...|0) ~,
\end   {equation} 
$ k_{i}=0,1,2 ~~ i,j=1,...n,~~ i\neq j $.
We see that there are four types of vertexes. A vertex with 
only one down field
corresponds to
points $3^{\prime}$ and $4^{\prime}$ on Fig.\ref{F3}c.
Vertex ${\bar \phi _{d}}\phi _{d}$  corresponds to point 6'
and vertex $\phi _{d}{\bar \phi _{d}}$ corresponds to point 5'. Vertex
${\bar \phi _{d}}\phi _{d}^{2}$  corresponds to points 1' and 2'.

An analytical expression which corresponds to contractions of leaves
comes out if we introduce the Boltzmannian field $\phi _{u}$
which has a decomposition (\ref {3}) with algebraic relations (\ref 
{5}),(\ref{6}).  Since upper and down lines  do not "observe" each 
others it is natural to assume that the corresponding fields are 
mutually commuted.

\begin{figure}
\begin{center}
\unitlength=0.50mm
\special{em:linewidth 0.8pt}
\linethickness{0.8pt}
\begin{picture}(193.00,188.12)
\put(85.00,114.00){\vector(0,1){23.00}}
\put(85.00,137.00){\vector(1,1){20.00}}
\put(105.00,157.00){\vector(-1,1){10.00}}
\put(95.00,167.00){\vector(1,1){10.00}}
\put(45.00,167.00){\vector(1,1){10.00}}
\put(45.00,167.00){\vector(-1,1){10.00}}
\put(35.00,177.00){\line(-1,2){1.00}}
\put(35.00,177.00){\line(1,2){1.00}}
\put(55.00,177.00){\line(-1,2){1.00}}
\put(55.00,177.00){\line(1,2){1.00}}
\put(95.00,167.00){\line(-1,2){1.00}}
\put(105.00,157.00){\line(1,1){2.00}}
\bezier{70}(27.00,187.00)(45.00,141.00)(63.00,187.00)
\bezier{80}(22.00,182.00)(85.00,89.00)(140.00,182.00)
\put(85.00,137.00){\vector(-4,3){40.00}}
\bezier{4}(22.00,182.00)(20.00,186.00)(19.00,188.00)
\put(32.00,85.00){\oval(10.00,10.00)[b]}
\put(56.00,85.00){\line(-1,2){1.00}}
\put(56.00,85.00){\line(1,2){1.00}}
\put(75.00,85.00){\line(-1,2){1.00}}
\put(75.00,85.00){\line(1,2){1.00}}
\put(117.00,85.00){\line(-1,2){1.00}}
\put(117.00,85.00){\line(1,2){1.00}}
\put(134.00,85.00){\line(0,1){2.00}}
\put(134.00,89.00){\line(0,1){2.00}}
\put(134.00,93.00){\line(0,1){2.00}}
\put(134.00,85.00){\line(0,1){2.00}}
\put(134.00,89.00){\line(0,1){2.00}}
\put(134.00,93.00){\line(0,1){2.00}}
\put(107.00,85.00){\line(0,1){2.00}}
\put(107.00,89.00){\line(0,1){2.00}}
\put(107.00,93.00){\line(0,1){2.00}}
\put(107.00,85.00){\line(0,1){2.00}}
\put(107.00,89.00){\line(0,1){2.00}}
\put(107.00,93.00){\line(0,1){2.00}}
\put(40.00,88.00){\makebox(0,0)[cc]{$^{ 1^{\prime}}$}}
\put(62.00,88.00){\makebox(0,0)[cc]{$^{ 3^{\prime}}$}}
\put(50.00,88.00){\makebox(0,0)[cc]{$^{ 2^{\prime}}$}}
\put(81.00,88.00){\makebox(0,0)[cc]{$^{ 4^{\prime}}$}}
\put(104.00,88.00){\makebox(0,0)[cc]{$^{ 6^{\prime}}$}}
\put(122.00,88.00){\makebox(0,0)[cc]{$^{ 7^{\prime}}$}}
\put(137.00,88.00){\makebox(0,0)[cc]{$^{ 5^{\prime}}$}}
\put(30.00,88.00){\makebox(0,0)[cc]{$^{ O^{\prime}_{1}}$}}
\put(85.00,142.00){\makebox(0,0)[cc]{$^{ 1}$}}
\put(90.00,118.00){\makebox(0,0)[cc]{$^{ O_{1}}$}}
\put(109.00,155.00){\makebox(0,0)[cc]{$^{ 5}$}}
\put(109.00,175.00){\makebox(0,0)[cc]{$^{ 7}$}}
\put(89.00,166.00){\makebox(0,0)[cc]{$^{ 6}$}}
\put(46.00,172.00){\makebox(0,0)[cc]{$^{ 2}$}}
\put(40.00,177.00){\makebox(0,0)[cc]{$^{ 3}$}}
\put(50.00,177.00){\makebox(0,0)[cc]{$^{ 4}$}}
\emline{56.00}{85.00}{1}{54.99}{87.08}{2}
\emline{54.99}{87.08}{3}{54.99}{87.08}{4}
\emline{54.00}{89.00}{5}{53.06}{90.95}{6}
\emline{52.00}{93.00}{7}{51.13}{95.09}{8}
\emline{56.00}{85.00}{9}{56.93}{87.08}{10}
\emline{58.00}{89.00}{11}{59.14}{90.95}{12}
\emline{60.00}{93.00}{13}{61.07}{95.09}{14}
\emline{75.00}{85.00}{15}{74.05}{87.08}{16}
\emline{73.00}{89.00}{17}{72.12}{90.95}{18}
\emline{71.00}{93.00}{19}{69.91}{95.09}{20}
\emline{75.00}{85.00}{21}{75.98}{87.08}{22}
\emline{77.00}{89.00}{23}{77.91}{90.95}{24}
\emline{79.00}{93.00}{25}{80.12}{95.09}{26}
\emline{117.00}{85.00}{27}{116.02}{87.08}{28}
\emline{115.00}{89.00}{29}{114.09}{90.95}{30}
\emline{113.00}{93.00}{31}{111.88}{95.09}{32}
\emline{117.00}{85.00}{33}{117.96}{87.08}{34}
\emline{119.00}{89.00}{35}{119.89}{90.95}{36}
\emline{121.00}{93.00}{37}{122.10}{95.09}{38}
\emline{150.00}{114.00}{39}{150.05}{115.86}{40}
\emline{150.00}{118.00}{41}{150.05}{120.04}{42}
\emline{150.00}{122.00}{43}{150.05}{123.83}{44}
\emline{150.00}{126.00}{45}{150.05}{128.00}{46}
\emline{150.00}{130.00}{47}{150.05}{129.90}{48}
\emline{150.00}{130.00}{49}{150.05}{132.18}{50}
\emline{150.00}{134.00}{51}{150.05}{135.97}{52}
\emline{150.00}{138.00}{53}{150.05}{140.14}{54}
\emline{150.00}{142.00}{55}{150.05}{143.94}{56}
\emline{150.00}{146.00}{57}{150.05}{148.11}{58}
\emline{150.00}{150.00}{59}{150.05}{151.90}{60}
\emline{150.00}{154.00}{61}{150.05}{156.08}{62}
\emline{150.00}{158.00}{63}{150.05}{159.87}{64}
\emline{150.00}{162.00}{65}{150.05}{164.04}{66}
\emline{150.00}{166.00}{67}{150.05}{165.94}{68}
\emline{150.00}{166.00}{69}{150.05}{167.84}{70}
\emline{150.00}{170.00}{71}{150.05}{172.01}{72}
\emline{150.00}{174.00}{73}{150.05}{176.18}{74}
\emline{150.00}{178.00}{75}{150.05}{178.08}{76}
\emline{150.00}{178.00}{77}{150.05}{179.97}{78}
\emline{150.00}{182.00}{79}{150.05}{184.15}{80}
\emline{150.00}{186.00}{81}{150.05}{187.94}{82}
\bezier{2}(140.00,182.00)(141.00,183.00)(144.00,188.00)
\emline{164.00}{177.00}{83}{162.95}{178.84}{84}
\emline{162.00}{181.00}{85}{161.05}{183.01}{86}
\emline{160.00}{185.00}{87}{159.16}{187.18}{88}
\emline{164.00}{177.00}{89}{164.85}{178.84}{90}
\emline{166.00}{181.00}{91}{167.12}{183.01}{92}
\emline{168.00}{185.00}{93}{169.02}{187.18}{94}
\emline{164.00}{114.00}{95}{164.09}{176.94}{96}
\put(156.00,118.00){\makebox(0,0)[cc]{$^{ O_{2}}$}}
\put(170.00,118.00){\makebox(0,0)[cc]{$^{ O_{3}}$}}
\emline{150.00}{85.00}{97}{150.00}{87.00}{98}
\emline{150.00}{89.00}{99}{150.00}{91.00}{100}
\emline{150.00}{93.00}{101}{150.00}{95.00}{102}
\emline{164.00}{85.00}{103}{163.00}{87.00}{104}
\emline{162.00}{89.00}{105}{161.00}{91.00}{106}
\emline{160.00}{93.00}{107}{159.00}{95.00}{108}
\emline{164.00}{85.00}{109}{165.00}{87.00}{110}
\emline{166.00}{89.00}{111}{167.00}{91.00}{112}
\emline{168.00}{93.00}{113}{169.00}{95.00}{114}
\put(17.00,88.00){\makebox(0,0)[cc]{$^{ O^{\prime}_{3}}$}}
\put(24.00,88.00){\makebox(0,0)[cc]{$^{ O^{\prime}_{2}}$}}
\put(35.30,176.91){\line(1,3){0.72}}
\put(54.86,176.91){\line(-1,2){0.95}}
\put(52.95,180.72){\line(-1,2){0.95}}
\put(51.04,184.54){\line(-1,2){0.95}}
\put(55.10,176.91){\line(1,2){0.95}}
\put(57.01,180.72){\line(1,2){0.95}}
\put(58.91,184.54){\line(1,2){0.95}}
\put(104.85,177.00){\line(-1,2){1.00}}
\put(104.85,177.00){\line(1,2){1.00}}
\put(99.85,177.00){\makebox(0,0)[cc]{$^{ 4}$}}
\put(104.71,176.91){\line(-1,2){0.95}}
\put(104.94,176.91){\line(1,2){0.95}}
\emline{55.10}{176.91}{115}{54.15}{178.81}{116}
\emline{53.19}{180.48}{117}{52.24}{182.39}{118}
\emline{51.28}{184.30}{119}{50.33}{186.21}{120}
\emline{55.10}{176.91}{121}{56.05}{178.81}{122}
\emline{56.77}{180.48}{123}{57.72}{182.39}{124}
\emline{58.68}{184.30}{125}{59.63}{186.21}{126}
\emline{94.93}{167.13}{127}{93.97}{169.04}{128}
\emline{93.02}{170.94}{129}{92.07}{172.85}{130}
\emline{91.11}{174.76}{131}{90.16}{176.67}{132}
\emline{89.20}{178.58}{133}{88.25}{180.48}{134}
\emline{87.30}{182.39}{135}{86.34}{184.30}{136}
\emline{85.39}{186.21}{137}{84.43}{188.12}{138}
\emline{104.94}{176.91}{139}{103.99}{178.81}{140}
\emline{103.04}{180.72}{141}{102.08}{182.63}{142}
\emline{101.13}{184.54}{143}{100.17}{186.45}{144}
\emline{104.94}{176.67}{145}{105.90}{178.81}{146}
\emline{106.85}{180.48}{147}{107.81}{182.39}{148}
\emline{108.76}{184.30}{149}{109.71}{186.21}{150}
\emline{105.18}{157.11}{151}{106.61}{159.02}{152}
\emline{108.04}{160.93}{153}{109.48}{162.84}{154}
\emline{110.91}{164.74}{155}{112.34}{166.65}{156}
\emline{113.77}{168.56}{157}{115.20}{170.47}{158}
\emline{116.63}{172.38}{159}{118.06}{174.28}{160}
\emline{119.49}{176.19}{161}{120.92}{178.10}{162}
\emline{122.35}{180.01}{163}{123.79}{181.91}{164}
\emline{125.22}{183.82}{165}{126.65}{185.73}{166}
\emline{179.00}{114.00}{167}{179.05}{115.86}{168}
\emline{179.00}{118.00}{169}{179.05}{120.04}{170}
\emline{179.00}{122.00}{171}{179.05}{123.83}{172}
\emline{179.00}{126.00}{173}{179.05}{128.00}{174}
\emline{179.00}{130.00}{175}{179.05}{129.90}{176}
\emline{179.00}{130.00}{177}{179.05}{132.18}{178}
\emline{179.00}{134.00}{179}{179.05}{135.97}{180}
\emline{179.00}{138.00}{181}{179.05}{140.14}{182}
\emline{179.00}{142.00}{183}{179.05}{143.94}{184}
\emline{179.00}{146.00}{185}{179.05}{148.11}{186}
\emline{179.00}{150.00}{187}{179.05}{151.90}{188}
\emline{179.00}{154.00}{189}{179.05}{156.08}{190}
\emline{179.00}{158.00}{191}{179.05}{159.87}{192}
\emline{179.00}{162.00}{193}{179.05}{164.04}{194}
\emline{179.00}{166.00}{195}{179.05}{165.94}{196}
\emline{179.00}{166.00}{197}{179.05}{167.84}{198}
\emline{179.00}{170.00}{199}{179.05}{172.01}{200}
\emline{179.00}{174.00}{201}{179.05}{176.18}{202}
\emline{179.00}{178.00}{203}{179.05}{178.08}{204}
\emline{179.00}{178.00}{205}{179.05}{179.97}{206}
\emline{179.00}{182.00}{207}{179.05}{184.15}{208}
\emline{179.00}{186.00}{209}{179.05}{187.94}{210}
\put(185.00,118.00){\makebox(0,0)[cc]{$^{ O_{4}}$}}
\put(7.00,88.00){\makebox(0,0)[cc]{$^{ O^{\prime}_{4}}$}}
\emline{35.00}{177.00}{211}{33.81}{179.05}{212}
\emline{33.00}{181.00}{213}{31.90}{182.86}{214}
\emline{31.00}{185.00}{215}{30.00}{187.14}{216}
\emline{35.00}{177.00}{217}{36.19}{179.05}{218}
\emline{37.00}{181.00}{219}{38.10}{182.86}{220}
\emline{39.00}{185.00}{221}{40.00}{187.14}{222}
\bezier{80}(16.00,114.00)(106.00,115.00)(193.00,114.00)
\bezier{80}(6.00,85.00)(96.00,86.00)(183.00,85.00)
\special{em:linewidth 0.4pt}
\linethickness{0.4pt}
\emline{85.00}{44.00}{223}{86.19}{48.10}{224}
\emline{86.00}{44.00}{225}{87.14}{48.10}{226}
\emline{85.00}{58.00}{227}{86.19}{61.90}{228}
\emline{86.00}{58.00}{229}{87.14}{61.90}{230}
\put(91.50,84.50){\oval(145.00,61.00)[b]}
\put(85.50,85.00){\oval(97.00,40.00)[b]}
\put(120.50,85.00){\oval(27.00,18.00)[b]}
\put(112.00,85.00){\oval(10.00,10.00)[b]}
\put(87.00,85.00){\oval(126.00,50.00)[b]}
\put(95.00,84.50){\oval(168.00,77.00)[b]}
\put(43.00,84.50){\oval(12.00,9.00)[b]}
\put(52.50,85.00){\oval(7.00,10.00)[b]}
\put(124.00,76.00){\vector(-1,0){5.00}}
\put(77.00,65.00){\vector(1,0){17.00}}
\put(62.00,84.50){\oval(26.00,25.00)[b]}
\put(164.00,135.00){\vector(0,1){0.09}}
\put(73.00,54.00){\vector(1,0){21.00}}
\put(30.00,80.00){\vector(1,0){2.08}}
\put(42.00,80.00){\vector(1,0){2.17}}
\put(52.00,80.00){\vector(1,0){0.92}}
\put(60.00,72.00){\vector(1,0){4.17}}
\put(111.00,80.00){\vector(1,0){2.00}}
\end{picture}
\end{center}
\vspace{-15mm}
\caption{ Trees contributing to four-point
Wightman's function}\label{F4} \end{figure}
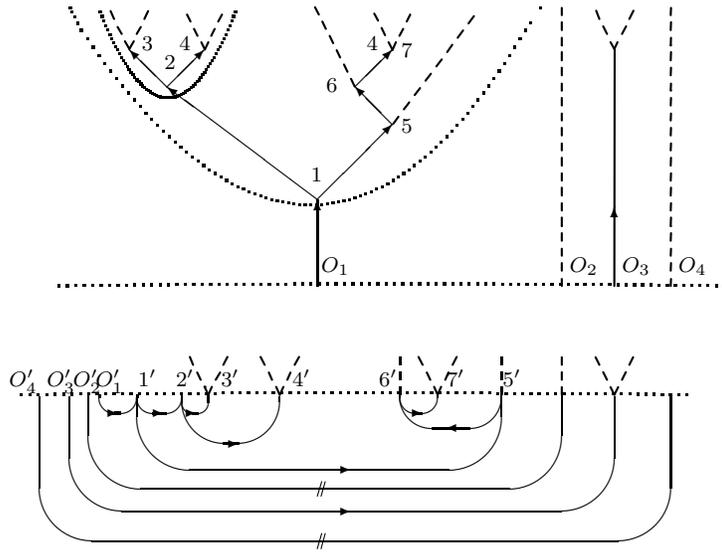

For some $\phi(x)$ we can take just a simplest
approximation $\phi(x)=\phi ^{in}(x)$ and therefore leaves may be
attached to some external points.
Note that  the ordering of this
field in respect to others $in$-fields attached to vertex lying on the
auxiliary line is essential.
To keep properly this ordering we add to ${\cal L}$ the
quadratic term
$((\Box + m^{2})\bar{\phi }_{d}(x)\phi _{u}$,
$\bar{\phi }_{d}(x)$ being contracted with the field
$\phi _{d}(x)$ after applying $\Box + m^{2}$  produces $\delta (x-x_{i})$
and gives $\phi _{u}$ which is in the right place (see Fig.\ref{F4}).
This remark completes the prove of the representation (\ref{1}).

In conclusion, 
we have shown that in the
large $N$ limit there exists the new "interaction" representation
for connected $SU(N)$-invariant Wightman's functions  of matrix fields
in term of scalar down
and up fields $\bar {\phi _{d}}(x)$, $\phi _{d}(x)$ and $\phi _{u}(x)$.
Note that the new interaction
representation contains  a rational function of the
interaction Lagrangian instead of
the exponential function in the standard interaction representation.
Selecting a part of terms in this interacting representation we get
so-called half-planar diagrams.
The Schwinger-Dyson equations  for the 2- and  4-point 
half-planar correlation
functions for this theory form a closed system of integral equations.
More rich set of terms in the representation (\ref{1})  reproduces
a parquet approximation (generalized ladder diagrams) in matrix
models \cite{AZpar}. By means of numerical calculations 
it has been  demonstrated
that in the large $N$ limit the parquet approximation for 0-dimensional
models gives a good
agreement with exact results.
\end {document}